\documentclass[onecolumn,prl]{revtex4}
\usepackage{bm}
\usepackage{epsfig}
\usepackage{amsmath}
\begin{document}
\title{Effective Thermal Diffusivity in Real Solid}
\author{Aniruddha Chakraborty \\
School of Basic Sciences, Indian Institute of Technology Mandi,\\
Mandi, Himachal Pradesh, 175001, India}
\date{\today }
\begin{abstract}
\noindent  The effective thermal diffusivity is evaluated for a two dimensional real solid and the real solid is modeled with periodic surface. The result contain the scale factor for the conformal transformation that flattens the surface. We find that the effective thermal diffusivity is always less than the local thermal diffusivity and this conclusion is independent of the model.
\end{abstract}
\maketitle
Temperature and heat flow are two important quantities in the understanding the heat conduction. When temperature distribution is not uniform at all points of the body, then heat flows in the direction of decreasing temperature. The Fourier equation [Eq.(1)] is in general used in understanding heat conduction which is valid only for a stationary homogeneous isotropic solid without any heat source. But in reality we have solids of different shapes, e.g. we can have a cylinder with nonuniform cylindricity. Now the aim of this paper is to understand the effect of this type of nonuniformity on the thermal diffusivity. As an example we consider a two dimensional solid modeled by a periodic surface and we look for an analytical expression for thermal diffusivity.
We start with Fourier equation of heat conduction \cite{Book},
\begin{equation}
\frac{1}{\alpha}\frac{\partial T(x,y,t)}{\partial t} = \bigtriangledown^2 T(x,y,t),
\end{equation}
where the constant $\alpha$ is called the thermal diffusivity of the medium, and is defined as
\begin{equation}
\alpha=\frac{k}{\rho c_p}.
\end{equation}
It combines three physical properties ($k$: thermal conductivity of the material, $\rho$: density of the material and $c_p$: specific heat) of the solid into a single constant. In Eq.(1) $T(x,y,t)$ is the temperature at location $(x,y)$ at time $t$. The differential equation of heat conduction will have numerous solutions unless a set of boundary conditions and an initial condition are fixed. We use standard no flux boundary condition
\begin{equation}
{\bf n}.\bigtriangledown T=0\;\;\; (on\;\; the \;\; boundary\;\; surface),
\end{equation}
where $n$ is the unit vector normal to the surface boundary. The shape of the solid is defined by 
\begin{equation}
-\infty < x < \infty \;\;\;\;\;\; and \;\;\;\;\; y_1(x) < y < y_2(x).
\end{equation}
$y_1(x)$ and $y_2(x)$ are the periodic functions of $x$ with periodicity $2 \pi$, so that
\begin{equation}
y_i(x+2 \pi)=y_i(x)\;\;\; i=1,2.
\end{equation}
The effective thermal diffusivity can be evaluated  by a similar procedure, as the one used in calculating the effective mass of an electron under a periodic potential. In the following we use the same procedure, we look for the eigenfunctions of the following equation
\begin{equation}
\bigtriangledown^2 \psi = - \lambda \psi
\end{equation}
with ${\bf n}.\bigtriangledown \psi=0$ at the boundary surface. As periodicity is involved in this problem, the eigenfunctions will have the popular Bloch form
\begin{equation}
\psi(x+ 2 \pi,y)=e^{2 \pi i k} \psi(x,y),
\end{equation}
where $k$ is the corresponding wave vector. For any value of $k$, there is a set of eigenvalues $\lambda_n(k)$. The smallest eigenvalue for any $k$ is denoted by $\lambda_0(k)$. So the effective thermal diffusivity $\alpha^*$ is given by the limit
\begin{equation}
\alpha^*=\alpha \lim_{k\rightarrow 0} \frac{\lambda_0 (k)}{k^2}.
\end{equation} 
Mathematically similar problems are already treated in a very interesting way by the use of complex variable technique \cite{Zwanzig,Weaver} using conformal mapping to flatten the surface boundaries \cite{Nehari}. We start with the conformal transformation from $w= u+ i v$ to $z=x+ i y$.
\begin{equation}
z= w + F(w),\;\;\;\;\; F(w + 2 \pi)=F(w).
\end{equation}
By proper choice of $F$, lines of constant $v$ in the $w$ planeare mapped into the surface boundaries in the $z$-plane. The period $0 < u < 2 \pi$ maps onto $0 < x < 2 \pi$. The problem in using conformal mapping is the difficulty in finding the function $F$ that flattens an arbitrary surface boundary. We follow the method used by Zwanzig \cite{Zwanzig}, chosing some trial functions $F$ and comparing surface boundary it leads to with the one we are considering in this problem. Transforming from $z$-plane to the $w$-plane we get the new Laplacian
\begin{equation}
\bigtriangledown^2 \rightarrow J \left(\frac{\partial^2}{\partial u^2}+\frac{\partial^2}{\partial v^2}\right),
\end{equation}
where $J$ is the Jacobian of the conformal mapping, and is given by
\begin{equation}
J= \left|  \frac{dw}{dz} \right|^2
\end{equation}
So the eigenvalue equation in $w$-plane becomes
\begin{equation}
\bigtriangledown^2 \psi = - \frac{\lambda}{J} \psi
\end{equation}
and $\frac{\partial \psi}{\partial v}=0$ at $v=v_1$ and $v=v_2$. The Block condition in $w$-plane will be
\begin{equation}
\psi(u,v)= e^{2 \pi i k } \psi(u,v).
\end{equation}
It is natural to introduce a new periodic function $\phi$,
\begin{equation}
\psi(u,v)= e^{i k u } \phi(u,v),
\end{equation}
where
\begin{equation}
\phi(u+ 2\pi,v)= \phi(u,v).
\end{equation}
The eigenvalue equation for $\phi$ is given by
\begin{equation}
\left(\frac{\partial^2}{\partial u^2}+\frac{\partial^2}{\partial v^2}\right)\phi+ 2 i k \frac{\partial \phi}{\partial u} - k^2 \phi= - \frac{\lambda}{J} \phi
\end{equation}
Now integrating the above equation over $u$ from $0$ to $2 \pi$ and over $v$ from $v_1$ to $v_2$, we get
\begin{equation}
\frac{\int_{v_1}^{v_2} dv \int_0^{2 \pi} du }{\int_{v_1}^{v_2} dv \int_0^{2 \pi} \frac{du}{J(u,v)} \phi (u,v)}= \lambda/k^2
\end{equation}
Now take the limit $k \rightarrow 0$, the resulting expression of thermal diffusivity involves only geometrical scale factor,
\begin{equation}
\frac{\alpha^*}{\alpha}=\frac{\int_{v_1}^{v_2} dv \int_0^{2 \pi} du \phi (u,v)}{\int_{v_1}^{v_2} dv \int_0^{2 \pi} \frac{du}{J(u,v)}}.
\end{equation}
In the following the shape of the solid is defined by its minimum and maximum widths i.e., $y_{min}=2[V-a sinh(V)]$ and $y_{max}=2[V+a sinh(V)]$. For a solid with symmetrical shape, where $v_2=-v_1=V$, the effective thermal diffusivity is given by
\begin{equation}
\frac{\alpha^*}{\alpha}= \left(1+\frac{1}{2} a^2 \frac{sinh(2V)}{2V}\right)^{-1}.
\end{equation}
If we keep the minimum width constant and allow the maximum width to approach infinity, under that condition the effective diffusivity can be expressed as
\begin{equation}
\frac{\alpha^*}{\alpha}= \frac{4}{y_{max}}
\end{equation} 
The effective thermal diffusivity is always smaller than the local thermal diffusivity. For a solid of rectangular shape, the thermal diffusivity is independent of the width of the solid. But any deviation from the rectangular shape provide extra regions, that the 'heat' can wander in before it can change its $x$-coordinate. So it takes more time to move in the $x$-direction, and hence thermal diffusivity is smaller. So our conclusion is true for any real solid and also for 3-dimensional solid. 
\par
{\it Its a pleasure to thank Dr. P. Anil Kishan for many interesting discussions}


\begin{thebibliography}{9}

\bibitem{Book} M. Necati Ozisik, {\it Boundary Value Problems of Heat Conduction}, (International Textbook company, Pannsylvania, USA, 1968).

\bibitem{Zwanzig} R. Zwanzig, Physica, {\bf 117A}, {\it 277} (1983).

\bibitem{Weaver} D. L. Weaver, Physica, {\bf 128A}, {\it 565} (1984).

\bibitem{Nehari} Z. Nehari, {\it Conformal Mapping}, (McGraw-Hill, New York, 1952).
 
\end{thebibliography}
\end{document}